# Low-loss silicon core fibre platform for mid-infrared nonlinear photonics

H. Ren[1], L. Shen[1,2,*], A. F. J. Runge[1,3], T. W. Hawkins[4], J. Ballato[4], U. Gibson[5,6] and A. C. Peacock[1]

[1] Optoelectronics Research Centre, University of Southampton, SO17 1BJ, UK

[2] Wuhan National Laboratory for Optoelectronics (WNLO), Huazhong University of Science and Technology, Wuhan 430074, China

[3] Currently with the Institute of Photonics and Optical Science (IPOS), School of Physics, University of Sydney, Australia

[4] Department of Materials Science and Engineering, Clemson University, Clemson, South Carolina 29634, USA

[5] Department of Physics, Norwegian University of Science and Technology (NTNU), N-7491 Trondheim, Norway

[6] Department of Applied Physics, KTH Royal Institute of Technology, Stockholm 10044, Sweden

*Corresponding author: L.Shen@soton.ac.uk

## Abstract

**Broadband mid-infrared light sources are highly desired for wide-ranging applications that span free-space communications to spectroscopy. In recent years, silicon has attracted great interest as a platform for nonlinear optical wavelength conversion in this region, owing to its low losses (linear and nonlinear) and high stability. However, most research in this area has made use of small core waveguides fabricated from the silicon-on-insulator platform, which suffer from high absorption losses of the silica cladding, limiting their ability to generate light beyond 3 μm. Here we design and demonstrate a compact silicon core, silica clad waveguide platform that has low losses across the entire silicon transparency window. The waveguides are fabricated from a silicon core fibre that has been tapered to engineer the mode properties to ensure efficient nonlinear propagation in the core with minimal interaction of the mid-infrared light with the cladding. These waveguides exhibit many benefits of fibre platforms, such as high coupling efficiency and power handling capabilities, allowing for the generation of mid-infrared supercontinuum spectra with high brightness and coherence spanning almost two octaves (1.6-5.3 μm).**

## Introduction

With wavelengths spanning from 2 to 20 μm, the mid-infrared (mid-IR) is an important spectral region, where strong molecular absorption bands and atmospheric transmission windows can be exploited for practical use in medicine, food production, imaging, environmental monitoring, and security.[1,2] For applications that require broad spectral bandwidths, such as spectroscopic sensing[3] and high resolution imaging,[4] supercontinuum (SC) sources based on extreme nonlinear phenomena have emerged as the most popular solution. To be effective, these sources must exhibit several key features including coherence, high brightness, robustness, stability and, for healthcare applications, be safe to handle.[5] Mid-IR SC spectra have been demonstrated in a range of material systems, in both fibre and planar platforms. To date, the broadest and brightest spectra have been demonstrated in fibre systems made from non-silica soft glasses (e.g., chalcogenides,[6] fluorides,[7] or tellurites[8]), primarily due to their capability to handle high powers. However, there are challenges to working with these materials as they are not as stable or robust as their silica counterparts and, in the case of the chalcogenides, they often contain toxic compounds. Alternatively, planar-based SC systems employing highly nonlinear Group IV semiconductor materials can avoid these issues, as well as offer advantages in terms of compactness and CMOS compatibility,[9] which are important considerations for the development of portable systems. In this case the trade-off is that these small core waveguides suffer from low power conversion efficiency, due both to the high on-chip coupling losses (typical 5-10 dB per facet) and propagation losses associated with increased core/cladding interactions.[10] As a result, the best demonstrations of SC generation in silicon-on-insulator (SOI) waveguides, the most common Group IV platform, have so far been limited to wavelengths less than 3.3 μm.[11,12] With a view towards extending the long wavelength edge, more complicated structures (i.e., suspended waveguides[13]) and material systems (e.g., silicon-on-sapphire[14-16] and silicon-germanium (SiGe)-on-silicon waveguides[17]), have been considered, though these come at a cost of increased fabrication and integration complexity.

Silicon core fibres (SCFs) represent an emerging platform that combines the benefits of the fibre geometry with the advantages of the semiconductor material systems. As these fibres are clad in silica, they are robust, stable, and fully compatible with standard fibre fabrication procedures, increasing device yield and reducing costs.[18] Until recently, one of the main limitations to producing SCFs using a fibre-drawing tower was obtaining the small, few-micrometre core sizes needed to observe efficient nonlinear processing. This is because, at the temperatures required to soften the silica cladding, the silicon core is molten, and high drawing speeds can lead to break-up of the core due to Rayleigh instabilities.[19] To overcome this hurdle, a post-draw tapering procedure was developed, which allows for more accurate control over the flow of the molten core, resulting in the smallest core crystalline silicon optical fibres with losses comparable to their planar waveguide counterparts.[20] This has allowed for the first demonstration of nonlinear propagation in the drawn SCF platform, extending from the telecoms band up to the mid-IR region.[21] Similar to the planar waveguides, SCFs have shown great potential for nonlinear processing in the mid-IR due to the reduced effects of two-photon absorption (TPA), but with two interesting advantages.

First, the larger SCF core sizes reduce the interaction with the silica cladding and increase the threshold for the higher order absorption mechanisms, allowing for higher power operation. Second, the SCFs can be directly spliced to other fibre components,[22] including the increasingly popular mid-IR fibre lasers,[23, 24] which opens a route to more efficient and elegantly packaged mid-IR systems.

In this work, we demonstrate a compact SCF platform capable of achieving a high brightness SC spectrum spanning almost two octaves, from the near infrared into the mid-IR. The fibre has been carefully designed in the shape of an asymmetric taper, which allows for improved coupling and minimal interaction of the long wavelength light with the silica cladding at the output. A femtosecond optical parametric oscillator (OPO) was used to pump the SCF at wavelengths around 3 µm, near the ZDW of the tapered waist, to generate a spectrally bright continuum covering 1.6-5.3 µm (∼3700 nm) with high coherence. The spectral broadening observed in the SCF represents the largest bandwidth generated in a silicon core/silica clad waveguide to date, with the long wavelength edge being pushed well beyond the silica absorption edge. The low propagation loss of our tapered SCF (<1 dB/cm in the mid-IR) enables a power conversion efficiency of ∼60%, which we believe is higher than any previous demonstration in a Group-IV-based waveguide and also comparable with non-silica fibre-based SC systems.[25] Furthermore, numerical simulations are used to show that the SCF platform could support SC generation extending out to 8 µm covering the entire transparency window of the silicon core.[26] This work provides a crucial step toward robust and practical all-fibre mid-IR broadband sources required for next generation healthcare and communication systems.

## Results

### Design of a low-loss mid-IR nonlinear SCF platform

The SCFs used in this work were fabricated using the molten core method (MCM) followed by tapering (as described in Ref. [27]); both procedures having been adapted from conventional silica fibre processing (see Methods). Significantly, as well as decreasing the fibre dimensions, the post-draw tapering also serves to improve the crystallinity of the silicon core to almost single-crystal-like quality, thus reducing the transmission losses.[28] Fig. 1a shows an optical microscope image of a tapered SCF with a constantly decreasing core diameter from 10 to 2 µm. As bulk silicon is transparent up to 8 µm, the linear propagation losses observed in silicon core/silica clad waveguides at the longer mid-IR wavelengths primarily originate from the overlap of the optical mode with the silica cladding glass, which is absorbing at these wavelengths. Thus, to achieve low propagation losses (<1 dB/cm) in this region, the cross-sectional geometry of the SCF is tailored to minimise this interaction for all generated wavelengths. As an illustration, Fig. 1b shows the linear losses at wavelengths beyond 3 µm for the SCFs with different diameters, as calculated via finite element method (FEM) simulations of the modal properties. In these simulations the real and imaginary parts of the complex refractive indices of silicon and silica are adapted from the literature.[29, 30] For example, a SCF with a core diameter of 2.8 µm exhibits low propagation losses (<1 dB/cm) for wavelengths up to ∼4.7 µm, and only a modest increase beyond this. This is as expected from the images of the fundamental mode for a fibre of this core size, shown in the inset of Fig. 1a, which confirms that the mode is well confined within the core over this wavelength range. Moreover, considering that typical lengths for silicon-based nonlinear devices are on the order of millimetres, the optical transmission in the micron-sized SCFs remains tolerable for even longer wavelengths, especially when compared with the submicron dimensions used for on-chip waveguides[10]. For example, an optical beam at a wavelength of 6 µm propagating through a length of 1 mm SCF with a 2.8 µm core diameter will only experience an extra 2 dB loss.

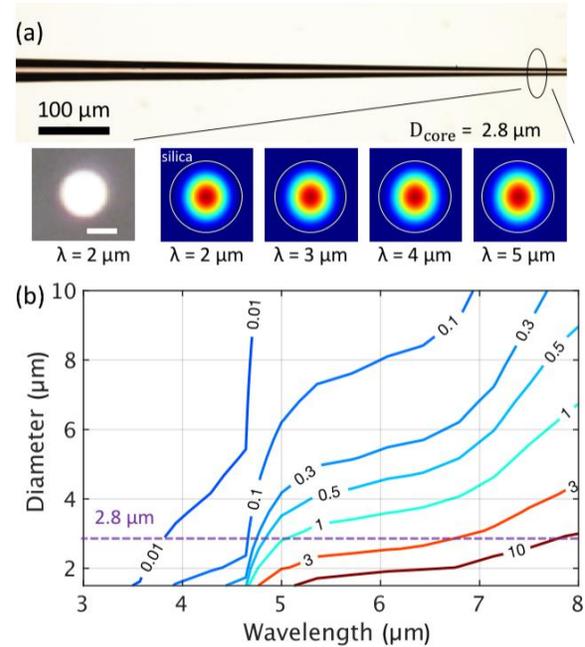

**Figure 1** (a) Microscope image of a tapered SCF with varying core diameter. Inset: measured output mode image (2 µm scale bar) and simulated mode profiles of a 2.8 µm core SCF at different wavelengths, as labelled. White lines highlight the boundary between silicon core and silica cladding. (b) Contour map showing simulated absorption losses for the SCFs in dB/mm as functions of wavelength and fibre core diameter.

As shown in Fig. 2a, the fibre is designed to have an asymmetrical profile, in which the large input core is slowly tapered down to ensure maximum coupling into the waist, minimising mode coupling and radiation loss, whilst a sharp up-taper is employed at the output to reduce the interaction with the lossy cladding. To achieve broadband mid-IR SC generation, the SCF was designed to be pumped at a wavelength around 3 µm. This pump wavelength was chosen because the nonlinear refractive of the silicon core is still relatively large, while the multiphoton absorption is modest as one approaches the three-photon absorption (3PA) edge.[31] Fig. 2b shows a contour map of the group velocity dispersion (GVD) parameter, $\beta_2$, as calculated via the well-known eigenvalue equation,[33] for the SCFs as functions of both wavelength and fibre core diameter. In order for the pump to access the anomalous dispersion region required for efficient SC generation,[32] the SCF only needs to be tapered down to have a core diameter less than 3.5 µm, which helps to ensure low propagation losses for wavelengths up to 6 µm, as illustrated in Fig. 1b. We note that although these micron-sized fibres support multiple guided modes, with careful coupling into the core, it is possible to propagate most of the light in the fundamental mode.[34] The exact profile of the fabricated tapered SCF is plotted in Fig. 2c. The fibre is gradually tapered down from a 10 µm core over the first 5.5 mm to reach a 2.8 µm diameter waist, of 1.5 mm length, followed by a 1 mm long inverse taper back up to a 10 µm core at the output. The waist length was chosen so that it was sufficiently long to induce a significant nonlinear phase shift, but without introducing prohibitively high losses for the longer wavelengths. The corresponding map of the zero-dispersion wavelength (ZDW) for this taper is shown in Fig. 2d, indicating the waist region is in the anomalous dispersion regime for the 3 µm pump. Interestingly, the varying dispersion profile of the taper provides another benefit in that it allows for multiple phase-matching conditions to be satisfied simultaneously, which can result in flatter and more coherent SC.[35, 36] By monitoring the low power transmission through the tapered SCFs, the average linear propagation losses are extracted to be within the range 0.2 to 3 dB/cm over the wavelength region 1.7-3.7 µm (see Supplementary Fig. 1), which are comparable to the lowest losses reported in on-chip silicon waveguides with similar dimensions.[15, 37]

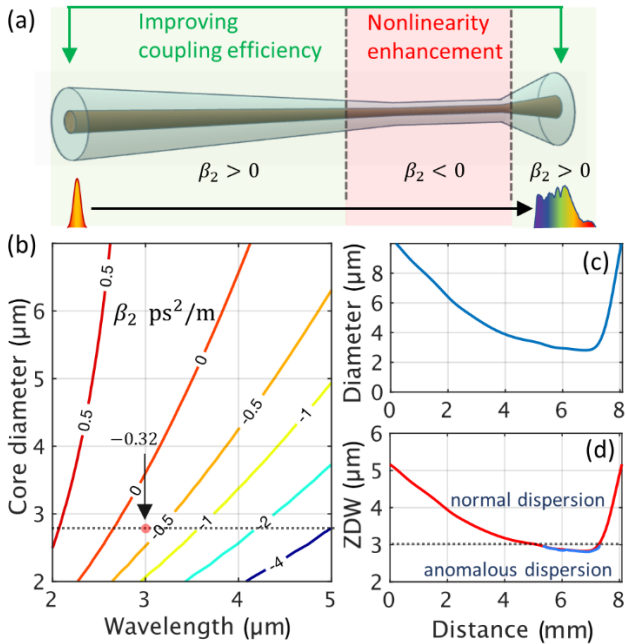

**Figure 2** (a) Schematic of the asymmetric taper design. (b) Contour map of the GVD parameters $\beta_2$ (ps$^2$/m) as functions of the core diameter and wavelength of the SCFs. (c) Profile of the core diameter variation in the fabricated tapered fibre. (d) ZDW profile of the fabricated tapered SCF.

## Experimental setup for supercontinuum measurements

The experimental setup for SC generation is shown in Fig. 3. A Ti:sapphire-pumped OPO with a ~100 fs (FWHM) pulse duration and an 80 MHz repetition rate was used as the pulse source. However, this source could easily be replaced with a femtosecond fibre laser that operates around the 3 µm regime,[38] to enable a compact and robust all-fibre SC source, similar to what has been proposed using a chalcogenide-based fibre taper in Ref. [35] The OPO was tuned to 3 µm and launched into the SCF via a black diamond objective lens L1 with a numerical aperture (NA) of 0.56, chosen to best match the fundamental mode size at the fibre input. The output light was collected by another black diamond objective lens L2 (NA=0.85). The coupling loss at the input fibre facet is estimated to be only 1.4 dB (excluding Fresnel reflections) at this pump wavelength. It is worth noting that this improved coupling efficiency, due to the taper structure, is significantly higher than previous demonstrations for on-chip mid-IR silicon waveguides,[11-17] which we attribute to better mode matching between the pump laser and the circularly symmetric fibre. CCD mid-IR cameras were employed at the input and output facets to facilitate coupling to the fundamental mode, as seen in the inset of Fig. 1a. Finally, the output spectra were recorded on a monochromator (Bentham TMc300).

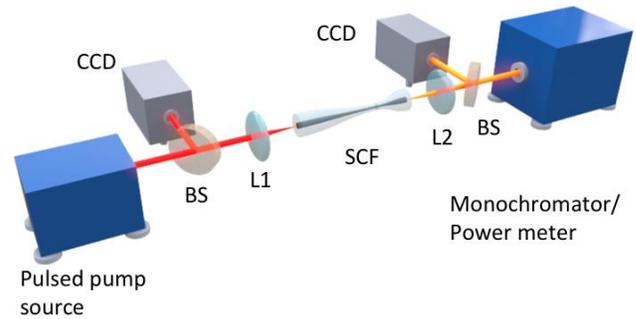

**Figure 3** Experimental setup of SC generation in a tapered SCF. The pulsed pump source is coupled into the tapered SCF with a lens (L1). The output light is collected with a second lens (L2) and measured by a power meter or a Monochromator via free space alignment. Beam splitters (BS) are used to direct light at the fibre input and output facets onto CCD cameras to monitor the coupling.

## Spectral evolution in tapered SCFs

The output SC spectra, obtained for coupled average pump powers, increasing ranging from 0.4 to 10.8 mW, are shown in Fig. 4a. For the highest power, the spectrum broadened significantly, spanning 1.72 octaves from 1.62 to 5.34 µm at the -40 dB level, for a maximum coupled peak power of only 1.19 kW, as shown in Table 1. The spectral broadening is dominated by self-phase modulation (SPM), followed by four wave-mixing (FWM) and dispersive wave (DW) emission, as evident from the evolution of the spectral profile and as labelled in the top spectrum. This evolution is similar to what has been previously reported in a deposited silicon fibre[39] and in SOI nanowires.[11, 12] However, the bandwidth obtained in the SCF has been significantly increased (almost twice as broad), extending well beyond the absorption edge of the silica cladding, thanks to the low-losses of the specially designed tapered structure. Moreover, the long wavelength signals (>5.5 µm) in the high power spectrum of Fig. 4a are still above noise level, however, we were not able to measure beyond this point due to the limited detector sensitivity. The offset of the noise floor at the short and long wavelength edges is due to different noise levels of the two detectors in the monochromator. Owing to the free space coupling arrangement, the atmospheric $CO_2$ absorption dip at 4.25 µm can be clearly observed in all spectra when the coupled input average power is above 6 mW. It is worth noting that a slightly reduced SC bandwidth, but with a longer wavelength edge (2.5-5.8 µm) has recently been generated in bulk silicon, but that this has required the use of a pump power that is three orders of magnitudes higher (MW level), owing to the lack of waveguide confinement.[40]

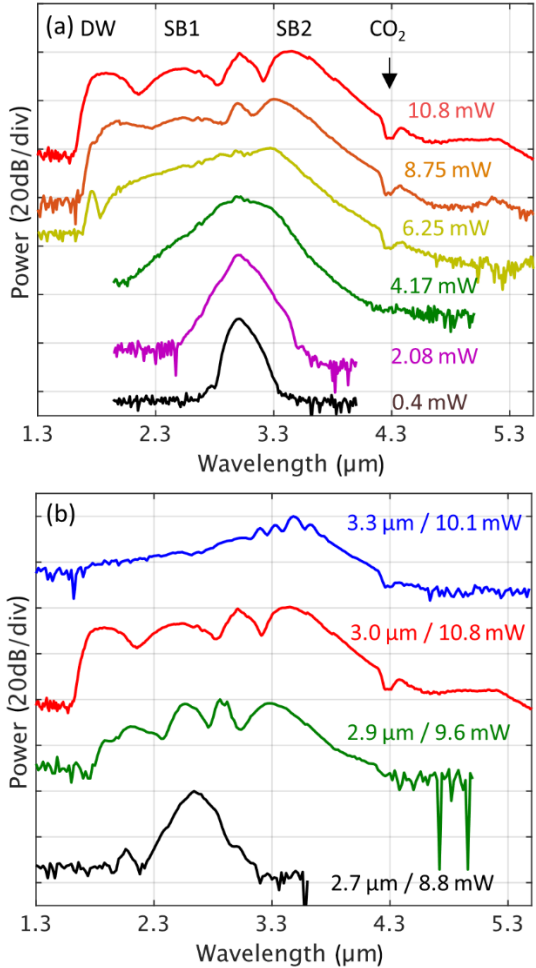

**Figure 4** (a) Experimental spectral broadening as a function of coupled average input power. The wavelength converted peaks associated with FWM (sidebands SB1 and SB2) and DW emission are labelled in the top spectrum. The black arrow shows the $CO_2$ absorption dip. (b) Comparison of SC generation with different pump wavelengths.

Fig. 4b shows the results from tuning the pump wavelength for four different wavelengths near 3 µm. For comparison, all the spectra have been recorded at their broadest bandwidth and the corresponding coupled average pump power is as labelled. As expected, the maximum spectral bandwidth is obtained at the optimum pump wavelength ($\lambda_p$ = 3 µm) where the tapered SCF exhibits low anomalous dispersion ($\beta_2$ = − 0.3 ps$^2$/m). In particular, when pumping in the normal dispersion regime at $\lambda_p$ = 2.7 µm the spectral broadening is mainly induced by SPM, so that it exhibits the smallest broadening of ~1100 nm. These results highlight the important role dispersion plays to enhance the broadening, with the broadest spectrum being obtained when the phase matching conditions for DW emission are met. By carefully designing the taper profiles for the desired pump wavelength the spectral broadening could be tailored for the wavelength region of interest. Nevertheless, all of the top three spectra in Fig. 4b exhibit more than an octave bandwidth at the -40 dB level, which shows that, owing to the range of core diameters accessible in the tapered SCF platform, it is more tolerant to variations in the fabricated dimensions than the straight waveguides where nanoscale precision is required for dispersion engineering.[41]

**Table 1. Supercontinuum results**

| Coupled Average Power (mW) | Coupled Peak Power (kW) | Spectral Range (µm) | Octaves |
|---|---|---|---|
| 10.8 | 1.19 | 1.62 - 5.34 | 1.72 |
| 8.75 | 0.96 | 1.66 - 4.51 | 1.44 |
| 6.25 | 0.69 | 1.68 - 4.45 | 1.41 |
| 4.17 | 0.46 | 1.94 - 3.96 | 1.03 |
| 2.08 | 0.23 | 2.48 - 3.59 | 0.53 |
| 0.41 | 0.045 | 2.62 - 3.37 | 0.36 |

Note: SC spectral range and octaves are taken with -40 dB bandwidth.

## Spectral brightness

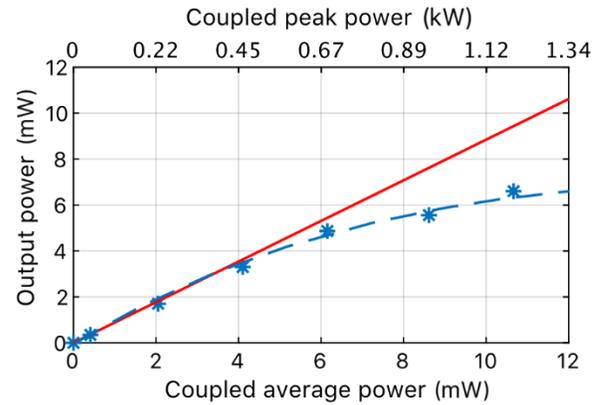

**Figure 5** Output SC power versus coupled power, average power (bottom) and peak power (top), for the tapered SCF. Red line indicates linear loss.

Fig. 5 shows the converted SC powers at the output facet of the tapered SCF for increasing coupled average (peak) power up to 10.8 mW (1.19 kW). The converted SC powers are back calculated by using the values measured on the detector and taking account the coupling loss at the output facet. A power conversion efficiency (the ratio of converted SC power $P_{out}$ to coupled average power $P_{in}$) as high as ~61% was achieved when the SC extends over a 3700 nm bandwidth at the -40 dB level. To the best of the authors' knowledge, this is the highest power conversion efficiency reported in any Group IV semiconductor platform, and is competitive even with the chalcogenide fibres.[17, 25] The converted SC power reached a value of 6.6 mW for 10.8 mW coupled average power, resulting in a spectrally bright SC with an average power spectral density of $PSD_{average}$= 0.002 mW/nm. The simulated SC powers (dashed curve), including multiphoton absorption (see Methods), show good agreement with the measurements. Unlike previous SC spectra obtained from similar coupled peak intensities (34.9 GW/cm$^2$),[15, 42] the total nonlinear loss introduced in the SCF was only 1-2 dB due to the short taper waist length. When combined with the high output coupling efficiency, a SC power of ~4 mW was obtained outside the fibre. This is a significant improvement over previous mid-IR SC demonstrations in silicon waveguides, which have been limited to out-coupled powers less than 1 mW.[15] As the pump power is far below the damage threshold of crystalline silicon, stable operation is expected and the tapered SCF used in this experiment exhibited no measurable change in transmission over several months of experiments.

## Numerical simulations and coherence properties

SC generation and its coherence properties can be simulated by numerically solving a generalised nonlinear Schrodinger equation (NLSE). As a starting point, Fig. 6a shows the simulated spectra as a function of pump power for comparison with the measured spectra in Fig. 4a. The wavelength dependent linear loss, second and third order dispersion ($\beta_2$, $\beta_3$), nonlinear refractive index ($n_2$), 3PA, free-carrier absorption (FCA) and dispersion (FCD) have been included in the model (see Methods). The simulated spectra show very good agreement with the measured results, both in terms of their bandwidths and the spectral features. For example, the -40 dB bandwidth of the simulated SC for 10.8 mW input power is 3600 nm (1.9 to 5.5 µm), which is almost identical to the experimental result shown in Table 1, and the three main spectral peaks (SB1, SB2 and DW) appear at similar positions. The slight difference between the simulations and experiments arises in the position of the DW, where there is a mismatch of ~0.2 µm. This discrepancy could be caused by uncertainties in the higher order dispersion values used in the simulations, principally due to the difficulty in precisely mapping the core diameter variations along the taper profile and waist. In addition, the discrete DW peaks on the short wavelength side are not visible in the experimental spectra because they are relatively weak and far away from the optimised coupling wavelength.

To better understand the role of the tapered profile, Fig. 6b and 6c show the simulated temporal and spectral evolution along the SCF for the maximum coupled peak power of 1.19 kW, clearly illustrating the complete spectral dynamics. The SC is generated through three steps. In the first section, between 0 and 4 mm of the fibre, the pump pulse is propagating in the normal dispersion regime and exhibits modest broadening due to SPM. As the core diameter continues to reduce, the pump pulses access the anomalous dispersion region and eventually it reaches the waist region at 5.5 mm. In this section, efficient phase matching occurs for the FWM process, which leads to pulse compression, break-up and eventually DW emission to produce a broadband SC spectrum. The final section incorporates the sharp, 1 mm long, up-taper to allow the SC light to be collected with minimal loss, during which there are no significant changes to the spectral components.

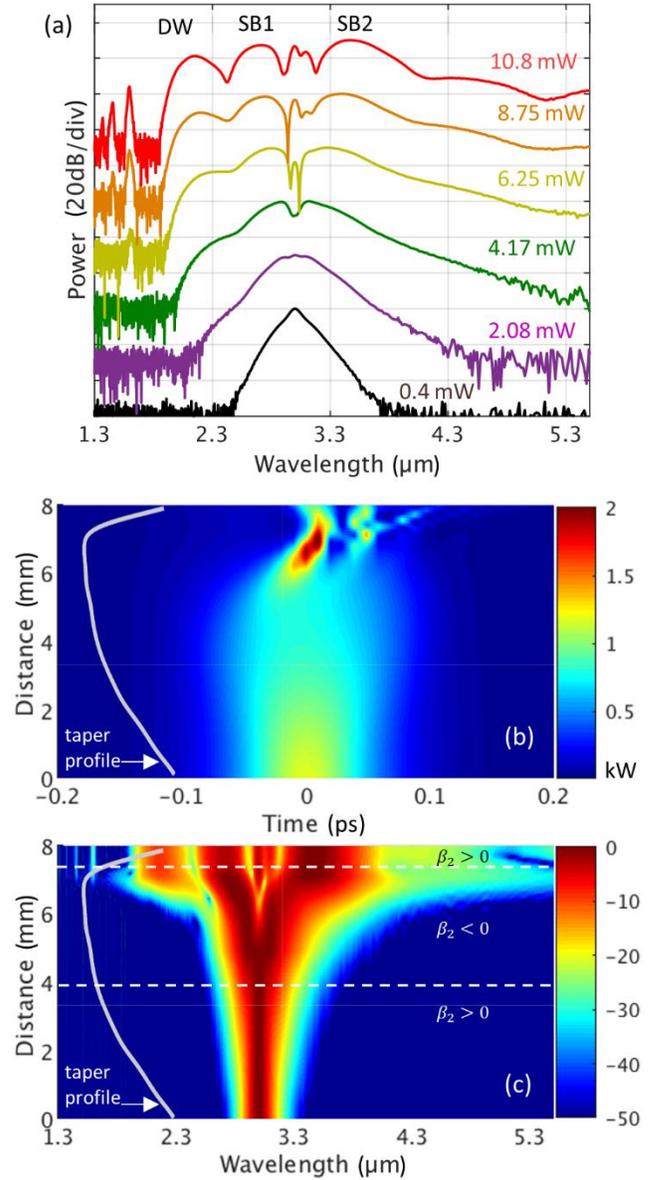

**Figure 6** (a) Numerical simulation results for SC spectra generated in a tapered SCF. (b) Simulated temporal evolution of the SC along the tapered fibre for the maximum coupled peak power of 1.19 kW. The colour bar shows the peak power in kW. (c) Simulated spectral evolution of the SC along the tapered fibre on a normalised logarithmic scale.

For applications such as spectroscopy, optical frequency comb metrology, and optical coherence tomography (OCT), it is important to ensure the generated SC can preserve the coherence of the pump laser. Unlike SC sources that make use of picosecond or longer pump pulses, where the broad spectral bandwidth is generated via amplification of background noise (modulation instability),[43] by relying largely on SPM and FWM, the SC generated in Fig. 4a is expected to have maintained good coherence, as reported in Refs. [11, 16]. The coherence of the SC was simulated via the method described in Ref. [11] using the mutual and self-coherence functions (see Supplementary 3) by including 200 different SC spectra. Each SC spectra is generated by incorporating 5% intensity variation together with quantum noise, included as one photon per mode with random phase, to the input pulse envelop.[43] As shown in Fig. 7, the SC generated in this tapered SCF is highly coherent (>0.9) and close to unity over its entire bandwidth, except for the region containing the low power discrete DWs on the short wavelength side. This is in good agreement with the predictions that SC spectra pumped by femtosecond pulses will largely preserve the coherence of the pump source.[44] The coherence could be further improved to approach unity if pumped by a more stable laser source (e.g., a mode-locked fibre laser) with negligible power fluctuations and intensity variation,[45, 46] a further compelling reason to move towards an all-fibre system.

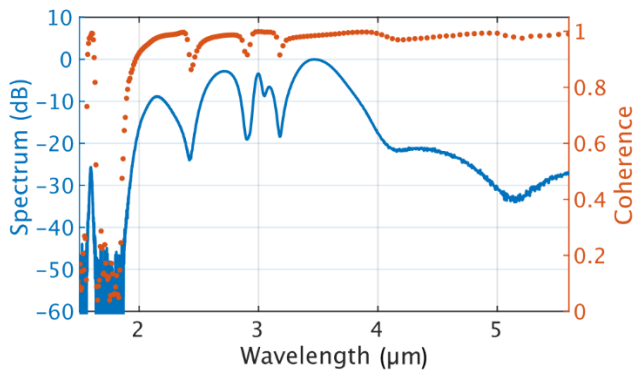

**Figure 7** Simulated coherence with quantum noise and ±5% power fluctuations shown in orange, together with the simulated SC spectra (blue) for the highest input power in Fig. 4a.

## Selective spectral improvement

The experimental results have shown that the tapered SCF platform is capable of generating SC spectra extending over 55% of the transparency window of the silicon core material. In order to access the remaining wavelength region, further optimization of the taper design is numerically investigated. To extend the blue edge of the SC, a straightforward solution is to reduce the taper waist and pump at shorter wavelengths, but this comes at the expense of cutting-off the longer wavelength light due to increased cladding absorption losses.[47, 48] As the emphasis here is on extending the long wavelength edge, a taper has been designed with a slightly larger waist diameter (3.1 μm) and a shorter overall length (details see Supplementary 4), as shown in Fig. 8b. Even with the high linear losses of the silica cladding for the longer wavelengths, the simulated spectrum shown in Fig. 8b can still reach beyond 8 μm at around the 20 dB level, when the fibre is pumped with an average input power of 27.2 mW (3 kW peak power). However, the modelling is limited to wavelengths up to 8 μm as the absorption in the silicon core increases substantially beyond this. As shown in Fig. 8a, the longer wavelength components (beyond 6 μm) are, in fact, generated in the up-tapered output region. Although this is beneficial for minimizing the long wavelength interaction with the cladding, it does mean that the broadening is very sensitive to variations in the output taper diameter as higher order dispersion plays a key role in the frequency conversion. Unfortunately, these experiments currently are limited by the accuracy of the tapering instrument used to produce such a SCF design. Nevertheless, this simulation indicates that the tapered SCF platform has potential to produce SC spectra that cover the entire silicon transparency window, even with an absorptive silica cladding. Further work could also consider replacing the core and cladding materials in a similar way to what is being explored in the planar platforms,[15, 17] though at the cost of introducing more fabrication complexity.

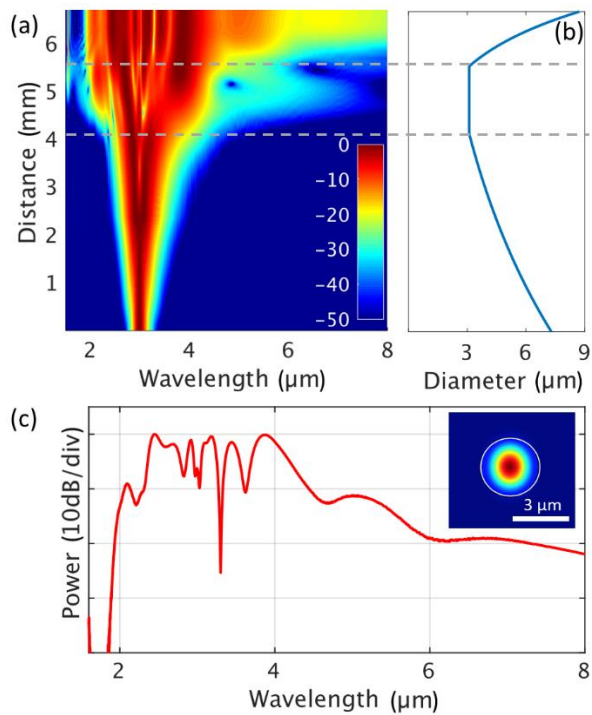

**Figure 8** (a) Simulated spectral broadening within a tapered SCF proposed for long wavelength generation. (b) Taper SCF profile. (c) Simulated SC output spectrum showing more than 2 octaves of broadening. Inset: mode image in the tapered waist at a wavelength of 8 μm.

## Discussion

A compact SCF platform has been demonstrated that achieves low-loss transmission across the mid-IR spectral regime. By exploiting a novel asymmetric taper design, a coherent SC spanning 1.74 octaves has been generated from 1.6-5.3 μm, which is the broadest SC reported in a silicon core/silica clad waveguide. The source also has high brightness and coherence, making it readily suitable for applications in mid-IR communications, spectroscopy and imaging. The experimental results, together with numerical simulations of an optimised profile, show that this tapered SCF platform could be used to generate SC spectra that cover the entire silicon transparency window up to 8 μm. Continued efforts to reduce loss and optimise the integration of this platform with other mid-IR fibre components will allow for the construction of robust, high power, and practical all-fibre SC-based mid-IR sources.

## Methods

### Fibre fabrication and post processing

The SCFs used in the experiments were fabricated using the MCM, whereby a silicon rod was sleeved inside a silica glass capillary to form a preform, which was subsequently drawn down into a fibre. The silica capillary was coated with a thin layer of calcium oxide (CaO), which forms an interface between the core and cladding during the drawing process. This layer plays an important role as it prevents dissolution of silica from the cladding into the silicon core and reduces the thermal strain arising from the high temperature processing. The core material used in this work was a high-resistance silicon rod (slightly phosphors doped, R>4800 Ω·cm) with negligible free carrier density. The fibre was drawn at a temperature of about 1950 °C with a drawing speed of approximately 25 m/min. The as-drawn silicon core materials are polycrystalline in nature, with crystalline grain sizes of a few hundreds of micrometres to millimetres in length. The as-drawn fibres have a silicon core diameter of ~20 μm with a ~140 μm silica cladding. The optical transmission losses of these fibres are around 10 dB/cm within the telecom band.

To improve the optical properties, the fibres can be thermally annealed or tapered using standard glass processing systems. Here, a Vytran GPX3400 was used to taper the fibres as this allows for control of the core dimensions post-draw. As the silicon core is completely molten during the tapering process, it is possible to use the drawing speed and temperature to control the cooling dynamics, which promotes large grain (centimetre length) crystal growth in the core.[28] The core/cladding interface of these fibres is also extremely smooth (root-mean-square roughness of ~0.7 nm), so that small core fibres (diameters <1 μm) can be obtained with minimal scattering and absorption losses.[28]

## Optical characterisation

Due to the variation of the core size along the length, the linear optical transmission losses of the tapered SCF were characterised by a single pass measurement. The fibre was mounted in a capillary tube and polished by using routine fibre preparation methods. The femtosecond OPO was employed as the light source to cover the full wavelength range of 1.7-3.7 μm. Thus, to ensure that nonlinear absorption was avoided, the average power of the injected light was kept below 100 μW. The light was launched into the core of the tapered SCF using the same configuration in SC measurements. The output of the fibre was imaged using a mid-IR camera to confirm that transmission occurred only through the silicon core. The coupling was optimised using a set of Thorlabs Nanomax stages. The power of the input and output beams were measured using two detectors; an InGaAs photodiode power sensor (Thorlabs S148C) for wavelengths <2.5 μm and a high resolution optical thermal power sensor (Thorlabs S320C) for wavelengths above this.

## Simulations

The Nonlinear Schrodinger equation (NLSE) was used to model pulse propagation along the tapered SCFs. A simplified model was employed, which includes the effects of 3PA, but not TPA, owing to the spectral position of the pump wavelength: [15]

$$\frac{\partial A}{\partial z} + \frac{i\beta_2}{2}\frac{\partial^2 A}{\partial t^2} - \frac{\beta_3}{6}\frac{\partial^3 A}{\partial t^3}$$
$$= ik_0 n_2 \frac{|A|^2}{A_{eff}} A - \frac{\beta_{3PA}|A|^4}{2A_{eff}^2} A - \frac{\sigma}{2}(1+i\mu)N_c A$$
$$- \frac{\alpha_l}{2} A$$

with

$$\frac{\partial N_c}{\partial t} = \frac{\beta_{3PA}}{3h\nu}\frac{|A|^6}{A_{eff}^3} - \frac{N_c}{\tau_c}.$$

Here A is the amplitude of the slow varying envelop of the optical pulse, z is the propagation distance, $\beta_2$ and $\beta_3$ are the second and third order dispersion of the fibre, $k_0$ is the propagation constant, $n_2$ is the nonlinear Kerr coefficient, and $A_{eff}$ is the effective mode area. $\alpha_l$ is the linear wavelength dependent loss. $\beta_{3PA}$ is the 3PA coefficient and $\sigma$, $\mu$, $N_c$, and $\tau_c$ are the FCA, FCD, 3PA induced free carrier density, and the free carrier life time, respectively. The values of these parameters can be found in the Supplementary Table 1. $h$ is Planck's constant and $v$ is the centre frequency of the pulse. This model only includes the second and third order dispersion terms since higher orders of dispersion typically contain large errors due to uncertainties in the wavelength dependent refractive index of the polycrystalline material and the precise core diameters. Raman effect is not included in our simulations as it has negligible impact on the generated SC and its cohenrence.[43] The equation was solved numerically using the well-known Split-step Fourier method.

## Acknowledgements

The authors acknowledge the following research funds: Engineering and Physical Sciences Research Council (EPSRC) (EP/P000940/1); National Natural Science Foundation of China (NSFC) (61705072); the Norwegian Research Council (262232); the J. E. Sirrine Foundation.

## Author contributions

A.C.P., L.S and H.R. conceived the research. T.W.H., J.B. and U.G. developed and fabricated the as drawn SCFs. H.R. and L.S. designed the taper profile and tapered the SCFs under A.C.P.'s guidance. H.R., A.F.J.R. and L.S. carried out the experiments and analysed the data. H.R., L.S. and A.C.P. wrote the manuscript; all authors contributed to the scientific discussions and revised the manuscript.

## Competing interests

The authors declare no competing financial interests.

# Supplementary information

## Supplementary 1 Transmission of the device

To characterise the transmission properties of the tapered SCF, the insertion loss of the device was measured for OPO wavelengths between 1.7-3.7 µm, using the lowest power setting to minimise the effect of nonlinear absorption. The transmission of the fibre was then estimated by excluding the 1.5 dB reflection from the facets and 2dB losses induced from the two lens. The large error bars are due to inconsistent coupling losses that vary for the different input wavelengths.

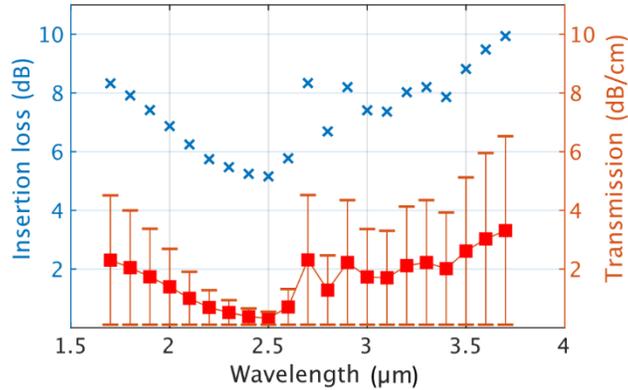

**Supplementary Figure 1:** Transmission properties of the tapered SCF at different wavelength.

## Supplementary 2 Simulation parameters

The optical parameters used in the numerical simulations are shown in Supplementary Table 1. The wavelength dependent loss was used in the simulation from the Supplementary Fig. 1. For input pulses with a duration of 100 fs, FCA and FCD were found to play a negligible role to the pulse shaping. The simulation uses z dependent dispersion and effective mode area profiles, as shown in Supplementary Fig. 2 and Supplementary Fig. 3, respectively.

**Supplementary Table 1.** SC simulation parameters

| Parameter | Symbol | Value (unit) |
|---|---|---|
| Central wavelength | $\lambda$ | 3 µm |
| Linear loss | $\alpha_l$ | Wavelength dependent losses shown in Supplementary Fig. 1 |
| Kerr effect coefficient | $n_2$ | $3.0 \times 10^{-18}$ m²/W from Ref. 1 |
| Pulse width | $T_{FWHM}$ | 100 fs (hyperbolic secant) |
| Carrier life time | $\tau$ | 10 ns |
| FCA parameter | $\sigma$ | $2.9 \times 10^{-21}$ m² from Ref. 2 |
| 3PA coefficient | $\beta_{3PA}$ | $2 \times 10^{-27}$ m³/W² from Ref. 1,3 |
| GVD parameter | $\beta_2$ | Shown in Supplementary Fig. 2 |
| 3rd order dispersion | $\beta_3$ | Shown in Supplementary Fig. 2 |
| Effective mode area | $A_{eff}$ | Shown in Supplementary Fig. 3 |

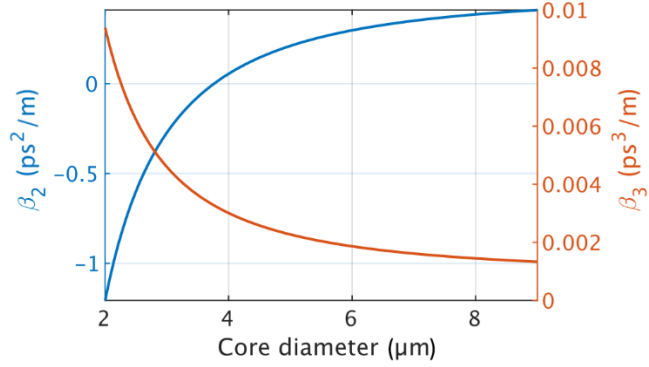

**Supplementary Figure 2:** Dispersion properties of the tapered SCF at the pump wavelength of 3 μm.

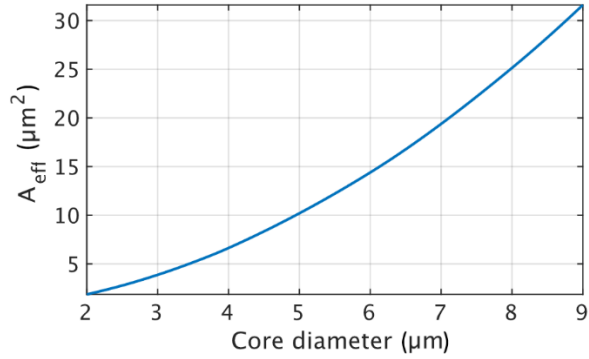

**Supplementary Figure 3:** Effective mode area of the tapered SCF at the pump wavelength of 3 μm.

## Supplementary 3 Coherence calculation

Coherence of the SC is calculated by the equation in Ref. 4:

$$|g_{12}(\lambda)| = \left| \frac{\langle A_1^*(\lambda) A_2(\lambda) \rangle}{\sqrt{\langle |A_1(\lambda)|^2 \rangle \langle |A_2(\lambda)|^2 \rangle}} \right|,$$

where $A_1$ and $A_2$ are the electric field amplitudes of two independent SC output. $g_{12}$ is the complex degree of first-order coherence and the angle bracket represents the ensemble average over all independently generated SC pairs. In this work, the coherence is calculated by applying 200 individual SC output with this equation. The laser source (OPO) used in this work has a 5% intensity noise with 30 dB level signal to noise ratio. To model the quantum noise, each of spectra was generated by incorporating one photon per mode with random phase on the input pulse envelop.

## Supplementary 4 Taper design for long wavelength

In order to generate longer wavelengths up to 8 μm for the same λ=3 μm pump source, the taper waist diameter is increased slightly (3.1 μm) and the overall fibre length shorten to 6.7 mm. This helps to minimize interaction of the mid-IR light with the cladding, reducing the transmission losses. In this case, the fibre is gradually tapered down from a 7.3 μm core over the first 4.1 mm to reach the diameter waist, of 1.4 mm length, followed by a 1.2 mm long inverse taper back up to a 9.3 μm core at the output. For an average input power of 27.2 mW (peak power of 3 kW), the generated SC could extend up to 8 μm at around the 20 dB level, as shown in the top spectrum of Supplementary Fig. 4b. From the temporal dynamics in Supplementary Fig. 4a, it is seen that the pulse compresses in the first 4 millimeters before breaking up in the waist region. Subsequently, the change of dispersion (anomolous to normal) in the up transition section of the taper enhances the pulse interactions, which further broadens the spectra. In this case, the dispersion profile needs to be precisely tailored to efficiently control the pulse interactions. Hence, this design is very sensitive to the ouput taper profile and has not yet been realised using our existing tapering equipment. Supplementary Fig. 4b shows the spectrum over different coupled peak powers.

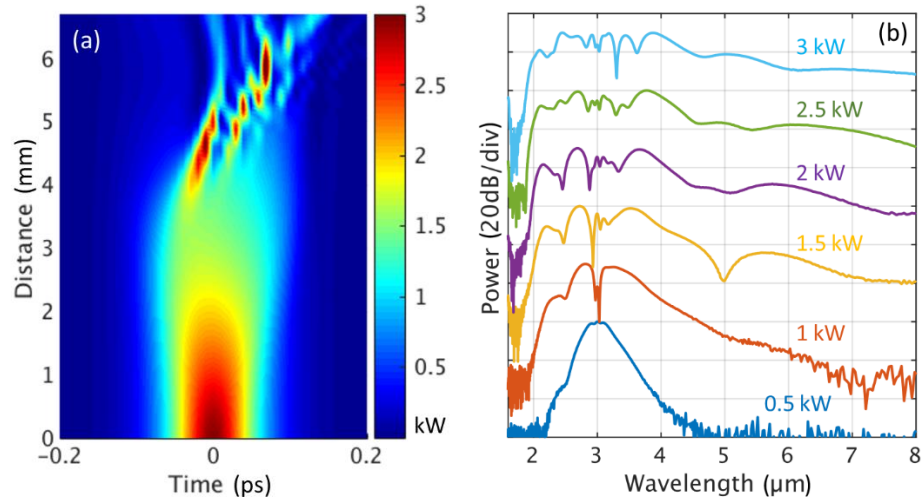

**Supplementary Figure 4:** Simulation results for the optimised taper structure. (a) Pulse evolution along the SCF in the time domain. (b) Simulated SC spectra with different coupled peak powers.

## Supplementary references: